\documentclass[a4paper]{article}
\usepackage[utf8]{inputenc}
\usepackage{natbib}
\usepackage[hidelinks]{hyperref}
\usepackage[margin=3.5cm]{geometry}
\usepackage{graphicx}
\usepackage{float}
\usepackage[margin=1cm]{caption}
\usepackage{amsmath}
\usepackage{amssymb}
\usepackage{appendix}
\author{Erik-Jan van Kesteren \& Tom Bergkamp}

\title{Bayesian Analysis of Formula One Race Results: Disentangling Driver Skill and Constructor Advantage}

\begin{document}
\maketitle
\begin{abstract}
    Successful performance in Formula One is determined by combination of both the driver's skill and race-car constructor advantage. This makes key performance questions in the sport difficult to answer. For example, who is the best Formula One driver, which is the best constructor, and what is their relative contribution to success? In this paper, we answer these questions based on data from the hybrid era in Formula One (2014 - 2021 seasons). We present a novel Bayesian multilevel rank-ordered logit regression method to model individual race finishing positions. We show that our modelling approach describes our data well, which allows for precise inferences about driver skill and constructor advantage. We conclude that Hamilton and Verstappen are the best drivers in the hybrid era, the top-three teams (Mercedes, Ferrari, and Red Bull) clearly outperform other constructors, and approximately 88\% of the variance in race results is explained by the constructor. We argue that this modelling approach may prove useful for sports beyond Formula One, as it creates performance ratings for independent components contributing to success.
\end{abstract}


\section{Introduction}
In most competitive sports with a large individual component (e.g., chess, athletics, swimming, or tennis) success is determined primarily by the relative skill (i.e., ability or talent) of the contestants. Official competitions in such sports naturally result in official rankings representing this skill level on an individual basis. Unlike these skill-based sports, competitive motor racing has an additional key factor contributing to success: the material, i.e., the race car or motor bike.  In Formula One in particular, the influence of the car on the results is considered to be substantial \citep{budzinski2020measuring}. In contrast to ``spec'' series, where cars have the same specifications and are built by the same constructor, Formula One cars are each built from the ground up by different constructors with differing levels of technological and financial resources. These resource gaps can lead to large differences in performance between cars, despite rules imposed to counteract such performance differences. Arguably, the presence of these large differences in constructor advantage have led to a single constructor (Mercedes) dominating the sport in the ``hybrid era'', from 2014 to 2020.

The dependence on materials reduces the relative influence of driver skill on success --- a race win is an entangled combination of both driver and constructor performance. Therefore, ranking drivers in terms of skill level by simply using competition race results is complex. Because of this problem, perennial questions such as ``who is the best Formula One driver?'', ``which constructor is the best?'' and ``is the driver or the constructor-team more important to success?'' are difficult to answer scientifically. These questions are persistent in the sport; the 2016 world champion Niko Rosberg famously posed that 80\% of success in Formula One can be attributed to the car and 20\% to the driver \citep{bol2020how}. In this article, we argue that it is now possible to answer  these questions due the widespread availability of race result data and the accessibility of advanced statistical methodology. We propose a novel Bayesian multilevel Beta regression model to answer three interrelated questions for the hybrid era in Formula One: (a) what is the relative influence of the driver and constructor on race results, (b) how do drivers rank in terms of skill level, and (c) how do constructors rank in terms of race car advantage?

Several attempts have been made to disentangle driver and constructor performance. \citet{eichenberger2009best} used linear regression with dummy variables and several covariates to estimate driver and constructor-year effects on race finishing position in the 1950-to-2006 Formula One seasons. The driver-specific effects on race outcomes were then used to compute a ranking of drivers' skill level. Additionally, the authors found that predictors for weather (wet vs. dry) and circuit type (street circuit vs. permanent circuit) were relevant additions to the model. One shortcoming of this study is the choice of outcome variable: because the number of contestants per race changed over seasons (and sometimes even within seasons), the interpretation of ``finishing position'' changed as well. This shortcoming was addressed by \citet{phillips2014uncovering}, who analysed Formula One race data from 1950-to-2013. Here, the effects of driver performance, constructor performance and season difficulty were estimated using an adjusted ``points scored'' outcome variable, based on the official points scoring system used in the Formula One between 1991 and 2002: 10 points for first place, 6 for second,  down to 1 point for sixth place. With this approach, the authors ranked drivers in terms of skill and concluded that Juan Manual Fangio was the best driver of all time.

As \citet{phillips2014uncovering} noted, there are two reasons that such models can differentiate driver from constructor effects: (a) throughout the history of Formula One, constructors have generally had two cars enter a race. Barring minor differences in individual races, these cars have the same performance, which allows for direct comparisons of a driver's skill level against their teammates. Furthermore, (b) drivers generally move to different constructor-teams throughout their career, meaning their teammates also change. This allows for simultaneous estimation of driver and constructor effects based on race results.

A disadvantage of the aforementioned studies is that they used models with many dummy variables (fixed effects) instead a multilevel (random effects) model. A multilevel approach is beneficial, as it makes the models tractable with fewer observations per driver and improves predictions for nested data \citep{gelman2006multilevel}. A similar argument was made by \citet{bell2016formula}, who used the same ``points scored'' outcome variable as \citet{phillips2014uncovering}, but used a multilevel (random-coefficients) linear model to determine the driver and constructor-year effects. Using this approach, the authors were the first to directly estimate a parameter for comparing driver skill and constructor advantage: they concluded that the constructor accounts for 86\% of the variance in points scored, and the driver for 14\%.

While the approach presented by \citet{bell2016formula} provides an answer to the research questions posed above (pre-2014), we here present several improvements which enable a more accurate insight into the current state of Formula One. First and foremost, we argue that the ``points scored'' variable leads to information loss: any result below sixth place leads to zero points, making it impossible to differentiate constructors and drivers who consistently finish below this threshold. Our approach, explained in section \ref{sec:model}, is to create a Rank-Ordered Logit model for the rank ordering of finish positions for each race. Second, we focus on the current hybrid era in Formula One, from 2014 -- 2021, using data from a publicly available resource \citep{newell2021ergast}. This focus enables us to closely inspect changes across these most recent seasons with comparable regulations. Third, we apply a Bayesian workflow for model development \citep{gelman2013bayesian}, visualisation \citep{gabry2019visualization}, and comparison \citep{vehtari2017practical}, which makes the parameters interpretable and clarifies the model's implications. Using our approach, the parameters for driver skill and constructor advantage in our model are directly interpretable as log-odds ratios of beating competitors, similar to Elo ratings in chess \citep[e.g.,][]{van2005psychometric}. These parameters (and their uncertainty intervals), can then be used to create rankings which provide clear insight into the relative performance of drivers and constructors in Formula One.

The paper is structured as follows. First, in Section \ref{sec:data} we describe the data source and processing steps we performed to obtain the predictors and the outcome of interest. Then, in Section \ref{sec:model} we introduce the proposed Bayesian multilevel regression model and its interpretation. In the same section, we also perform model selection and model checking to validate the estimation procedure. In Section \ref{sec:results} we perform inference for the 2014-to-2021 Formula One seasons to answer the research questions surrounding driver skill and constructor advantage. This includes a driver ranking for the 2021 season and investigation of a counterfactual statement based on the estimated model: would Hamilton in an Alfa Romeo beat Räikkönen in a Mercedes? Last, in Section \ref{sec:discussion} we place our contributions in context, discuss its implications, and  provide suggestions for future work. All analysis scripts and pre-processed data are openly available in the supplementary material at \texttt{\href{DOI:10.5281/zenodo.7632045}{https://doi.org/10.5281/zenodo.7632045}} \citep{erik_jan_van_kesteren_2023_7632045}.

\section{Data processing}
\label{sec:data}
We collected race results (driver id, constructor id, season (year), race number, finishing position and status) for the 2014-to-2021 Formula One seasons from the dataset behind the Ergast motorsports API \citep{newell2021ergast}. This resulted in a dataset of 160 races, with 51 unique drivers and 19 distinct constructors.
In addition, we performed a data enrichment step by scraping and parsing further race information from Wikipedia. In this step, two predictors were added to the dataset which were previously found to be relevant in the work by \citet{eichenberger2009best} and \citet{bell2016formula}. We constructed a variable indicating whether the race was wet or dry, and we also collected information about the circuit type (street circuit or permanent circuit). As the information was not complete, especially for weather type in the 2018 season, we manually completed this data using publicly available race summaries. In total, in the period of interest, there were 143 dry races and 17 wet races.

For the main analysis, non-finishers were removed from the analysis, removing 590 rows (from the original 3267 rows) in the dataset. Thus, we only examined finished races for each driver: accidents and other reasons for non-finishing or non-starting can be due to any number of reasons, which adds noise and complexity to the outcome of interest. This removal of non-finishes has implications for the interpretation of the model, which we discuss in section \ref{sec:model}. In Appendix \ref{app:finish} we perform a sensitivity analysis for the effects of dealing with finishing in different ways. Note that instead of the approach we take, it is also possible to include non-finishing in the analysis by creating a larger, joint model of non-finishes and race results conditional on finishing \citep[e.g.,][]{ingram2021first}.

\begin{figure}
    \centering
    \includegraphics[width=0.8\linewidth]{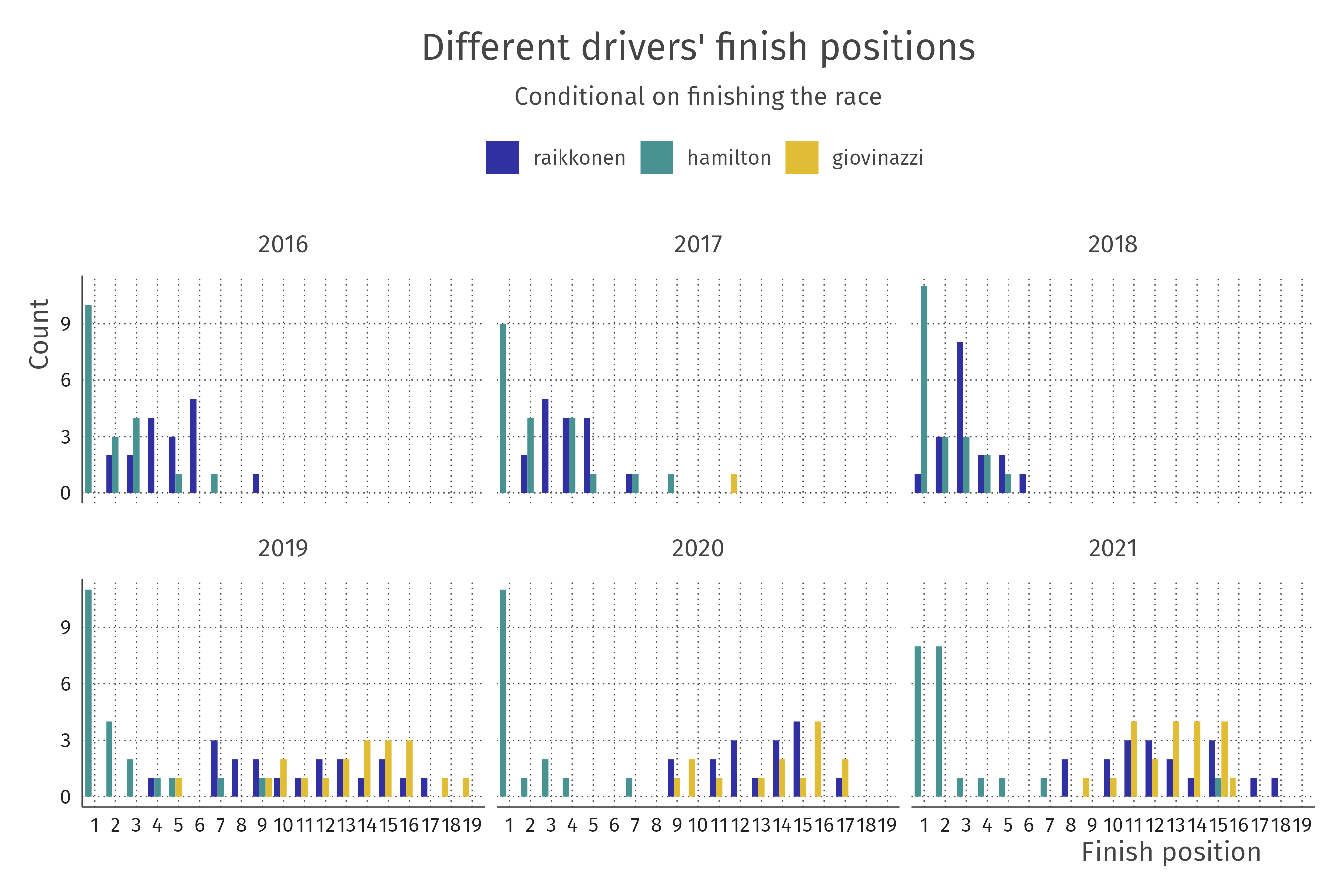}
    \caption{Finish positions for Räikkönen, Hamilton, and Giovinazzi for the seasons 2015 -- 2020. Räikkönen's move from Ferrari to Alfa Romeo in 2019 is clearly visible in the finish positions. At Alfa Romeo, Räikkönen became teammates with Giovinazzi, who he on average outperforms slightly.}
    \label{fig:prop}
\end{figure}

A visual display of the finish positions for three drivers (Räikkönen, Hamilton, and Giovinazzi) in the 2015-2020 seasons is shown in Figure \ref{fig:prop}. This forms the basis of the outcome variable we construct, which is the per-race ranking of competitors. By considering only race ranking as our outcome, we disregard the starting position of the competitor, which is determined in a qualifying session before the race. Thus, any conclusions about relative performance of competitors necessarily includes both race performance and qualifying performance: for all of our models, a race-winning performance is no different if the competitor starts the race in first or last position. In the next section, we explain in detail how we model this outcome variable.

\section{Bayesian multilevel rank-ordered logit model}
\label{sec:model}
In this section, we describe in detail the process used to model the finish position as a function of driver skill, yearly driver form, constructor advantage, and yearly constructor form. All models in this paper were estimated using the software package \texttt{Stan} \citep{carpenter2017stan} with reasonable default priors for all parameter types (see section 2.1 of \citet{burkner2017brms} for more details).\\

\subsection{Basic model specification and parameter interpretation}

For each race $r$, assume we have a set of competitors $\mathcal{C}_r$ which compete for the win. The outcome variable in our model is a vector representing a ranking of these competitors, $\boldsymbol{y}_{r}$. In our generative model, we assume that this ranking follows a Rank-Ordered Logit (ROL) distribution, based on a vector of latent abilities of the competitors in that race $\boldsymbol{\vartheta}_r$:
\begin{align}
    \boldsymbol{y}_{r} &\sim {\rm RankOrderedLogit}(\boldsymbol{\vartheta}_{r})
    \label{eq:rolmodel}
\end{align}

The precise definition of the ROL distribution is explained in detail in \citet{glickman2015stochastic}, and in our extended model definition in Appendix \ref{app:model}. Here, note that each element of $\boldsymbol{\vartheta}_{r}$ represents the latent ability of each competitor in the race:
\begin{align}
    \boldsymbol{\vartheta}_{r} &= \{\vartheta_{c} \, | \, c \in \mathcal{C}_r\}
\end{align}

A competitor $c \in \mathcal{C}_r$ is defined as a pairing of a driver $d$ and constructor (or team) $t$, in a particular season (or year) $s$. Thus, we represent the skill of each competitor as the skill of the driver-constructor pairing in a season: $\vartheta_{c} = \theta_{dts}$. In our model, this skill is a sum of the average driver skill $\theta_d$, the driver's seasonal form $\theta_{ds}$, the constructor's average advantage $\theta_{t}$ and the constructor's seasonal form $\theta_{ts}$. This results in the following cross-classified multilevel model for each competitor's latent ability:
\begin{align}
    \vartheta_{c} = \theta_{dts} &= \theta_d + \theta_{ds} + \theta_t + \theta_{ts} \nonumber \\
    \theta_d &\sim \mathcal{N}(0, \sigma^2_d) \nonumber \\
    \theta_{ds} &\sim \mathcal{N}(0, \sigma^2_{ds}) \nonumber \\
    \theta_t &\sim \mathcal{N}(0, \sigma^2_t) \nonumber \\
    \theta_{ts} &\sim \mathcal{N}(0, \sigma^2_{ts})
    \label{eq:model}
\end{align}



This model form leads to a specific, natural interpretation for the parameters $\theta$. The logit link function in the ROL likelihood, in combination with the omission of an overall intercept, ensures that the (hypothetical) average driver at an average constructor with an average seasonal form will on average have $\theta_{dts} = 0$, which translates into a probability of 0.5 of beating a randomly selected other competitor. Then, $\theta_d$ represents the mean driver skill as a log-odds ratio; e.g., if $\theta_d = 0.3$, this means (ceteris paribus) that the probability of finishing in front of the other driver is $1/(1+e^{-0.3}) \approx 0.57$. This parameter represents a deviation from the average driver, so negative values mean worse than average skill, and positive values mean better than average skill. We also include the seasonal driver form parameter $\theta_{ds}$, which represents yearly deviations from this long-term average driver skill. 

A similar interpretation holds for $\theta_t$, which indicates the long-term average constructor advantage. Constructors with positive values on this parameter tend to produce cars which are better than average, and negative values indicate cars which are worse on average. We also include the seasonal constructor form parameter $\theta_{ts}$, which represents yearly deviations from this long-term average team advantage. In Appendix \ref{app:dynamic} we compare this parameterization to two other ways of dealing with time-varying $\theta$ parameters. For more detailed parameter interpretations and conclusions, see Section \ref{sec:results}. 

Note that the latent skill are assumed to be stable within seasons. This means that for races within the same season $s$ with the same competitors $\mathcal{C}_r$ we assume independent and identically distributed rankings $\boldsymbol{y}_{r}$. Note also that with this model formulation we implicitly assume no correlation between the random intercepts for driver and car constructor; there are no interactions at all between driver skill and team advantage. This means that a driver's skill is independent of the team advantage, i.e., the driver skill does not change when the driver moves to a different constructor.

\subsection{Extending the basic model}
Previous work has shown that several predictors may change the race results \citep{bell2016formula}. The first extension we make reflects the knowledge that wet races are different from dry races. Wet races require a specific set of skills, which rely less on the car and more on the driver. Like \citet{bell2016formula}, we represent this knowledge by splitting the driver average skill parameter into a random intercept parameter $\gamma_{0d}$ and a random slope parameter $\gamma_{1d}$ as in Equation \ref{eq:wetdry}:

\begin{equation}
    \theta_d = \gamma_{0d} + \gamma_{1d} \cdot \text{wet\_race}
    \label{eq:wetdry}
\end{equation}

where wet\_race is an indicator (dummy) variable with a 1 if the race was wet and 0 if the race was dry (see Section \ref{sec:data} for details). The driver average skill in dry races is then $\gamma_{0d}$, and in wet races it is $\gamma_{0d} + \gamma_{1d}$.

The second extension we make reflects the knowledge that different constructors have different car philosophies. Theoretically, high-downforce concept cars (e.g., Red Bull cars in the hybrid era) are relatively better suited to narrow, curvy street circuits such as the famous Monaco circuit. However, this advantage disappears on fast, permanent circuits such as Monza (Italy). Therefore, we add a random slope to the constructor advantage parameter, splitting it up as in Equation \ref{eq:streetpermanent}: 

\begin{equation}
    \theta_t = \gamma_{0t} + \gamma_{1t} \cdot \text{permanent\_circuit}
    \label{eq:streetpermanent}
\end{equation}

where permanent\_circuit is an indicator (dummy) variable with a 1 if the race was on a permanent circuit and 0 if the race was on a street circuit (see Section \ref{sec:data} for details).

With these two extensions, four models are possible: (a) the basic model, (b) a weather model, (c) a circuit type model, and (d) a weather and circuit type model. In the next subsection, we compare these models to select our final model, on which we perform inference.

\subsection{Model selection}
We used efficient leave-one-out cross-validation \citep[LOO,][]{vehtari2017practical} to compare the four possible models. In short, LOO uses the samples from the posterior to compute the expected log posterior density (ELPD) for each model, which is an alternative to the standard information criteria in Bayesian model comparison such as the Bayes Factor (marginal density) or DIC. For the four tested models, the results are shown in Table \ref{tab:loo}.

\begin{table}[ht]
\centering
\begin{tabular}{rcccc}
  \hline
  & $ELPD$ & $SE_{ELPD}$ & $\Delta$ & $SE_{\Delta}$ \\ 
  \hline
  Circuit            & -3992.29 & 84.11 &       &      \\ 
  Basic              & -3993.36 & 83.72 & -1.07 & 6.35 \\ 
  Circuit + Weather  & -3993.66 & 84.31 & -1.37 & 1.94 \\ 
  Weather            & -3995.55 & 84.09 & -3.25 & 6.22 \\ 
  \hline
\end{tabular}
\caption{Model comparison results showing the expected log posterior density (ELPD), its standard error, the difference between each model and the best model, and the standard error of this difference. Note: a higher ELPD indicates a better fit.}
\label{tab:loo}
\end{table}

The circuit model is shown to be the best model in terms of ELPD, but the differences are small compared to the standard errors of these differences. This indicates that in terms of out-of-sample predictive performance, the models perform similarly. To maintain parsimony in both modelling and interpretation, we decide use the basic model for assessment, inference, and prediction in the following sections.

\subsection{Model assessment}

Posterior samples were obtained via Hamiltonian Monte Carlo sampling with 8 chains of 1250 samples each after 1000 burn-in iterations. The effective sample size for all parameters was higher than 2500, and their R-hat value was smaller than 1.01, indicating adequate convergence. Trace plots for the main parameters of the model are shown in Appendix \ref{app:trace}.

\begin{figure}[H]
    \centering
    \includegraphics[width=0.95\linewidth]{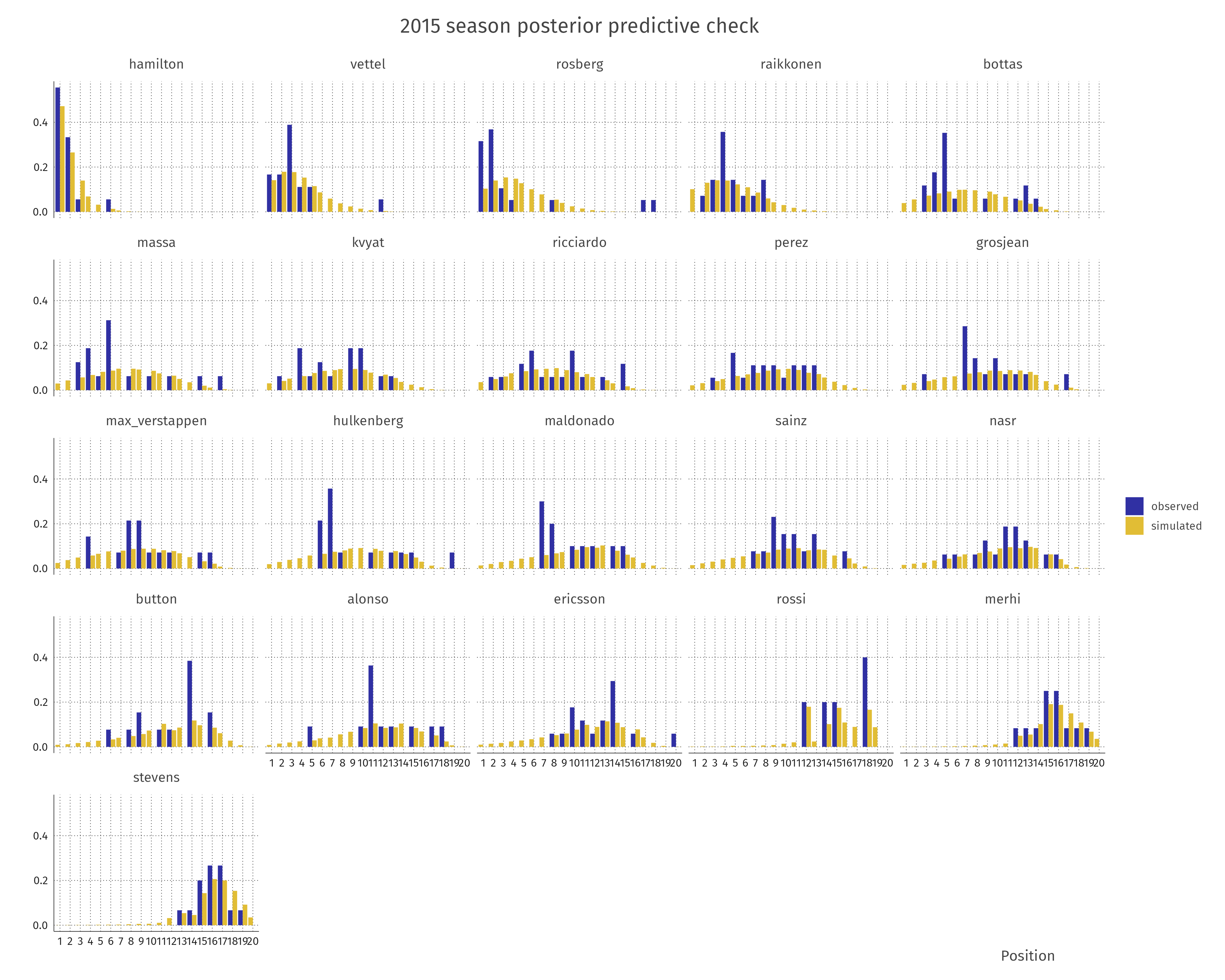}
    \caption{Posterior predictive check for driver finishing positions in the 2015 season.}
    \label{fig:ppc1}
\end{figure}

\begin{figure}[H]
    \centering
    \includegraphics[width=0.95\linewidth]{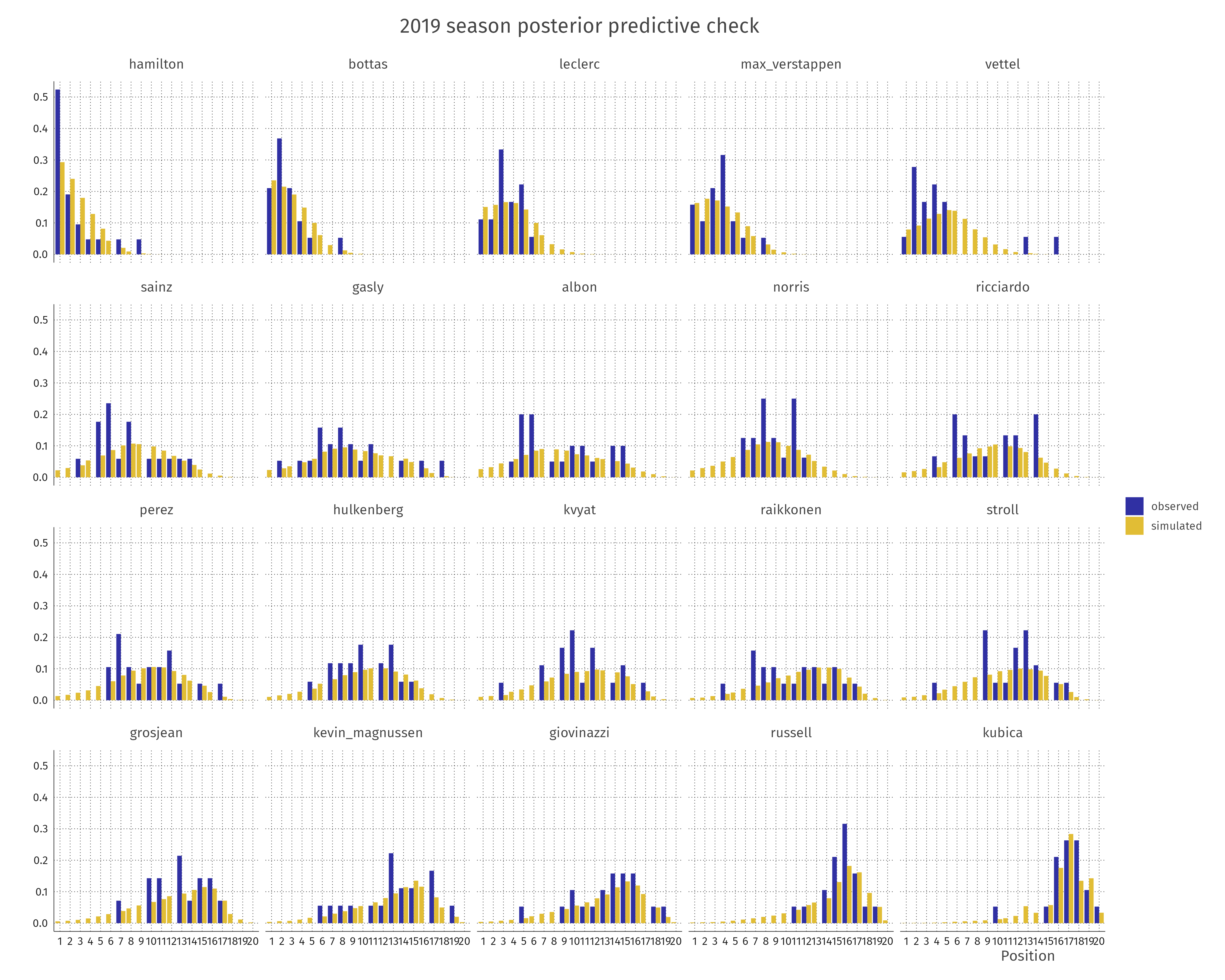}
    \caption{Posterior predictive check for driver finishing positions in the 2019 season.}
    \label{fig:ppc2}
\end{figure}

Posterior predictive checks (PPCs) are a vital part of the Bayesian workflow \citep{gabry2019visualization}. In a visual PPC, simulated data $\Tilde{y}$ from the posterior predictive distribution is compared to the observed data $y$. If $\Tilde{y}$ approximates $y$ well, then the model captures the outcome well. For our model, we performed two PPCs: one for the 2015 season (early hybrid era) and one for the 2019 season (late hybrid era). The results are shown in Figures \ref{fig:ppc1} and \ref{fig:ppc2}.


The plots show satisfactory recovery of individual performances per driver-constructor-season combination. It is noteworthy that for some drivers, the observed data show bimodality (e.g., Vettel in 2019 or Bottas in 2015) while the model cannot capture these patterns. In these cases, the averages in the observed and simulated data still seem to align closely. In general, we conclude that the model fits the observed data well. In the next section, we use the model to perform parameter inference.

\section{Results}
\label{sec:results}
In this section, we perform inference with the model that resulted from the modelling procedure described in Section \ref{sec:model}. To narrow down our inference efforts, we focus on a subset of the drivers and teams competing in the 2021 season. The results for all drivers, teams, and seasons are available in the supplementary material \citep{erik_jan_van_kesteren_2023_7632045}.

\subsection{Driver skill}
\begin{figure}[H]
    \centering
    \includegraphics[width=0.8\textwidth]{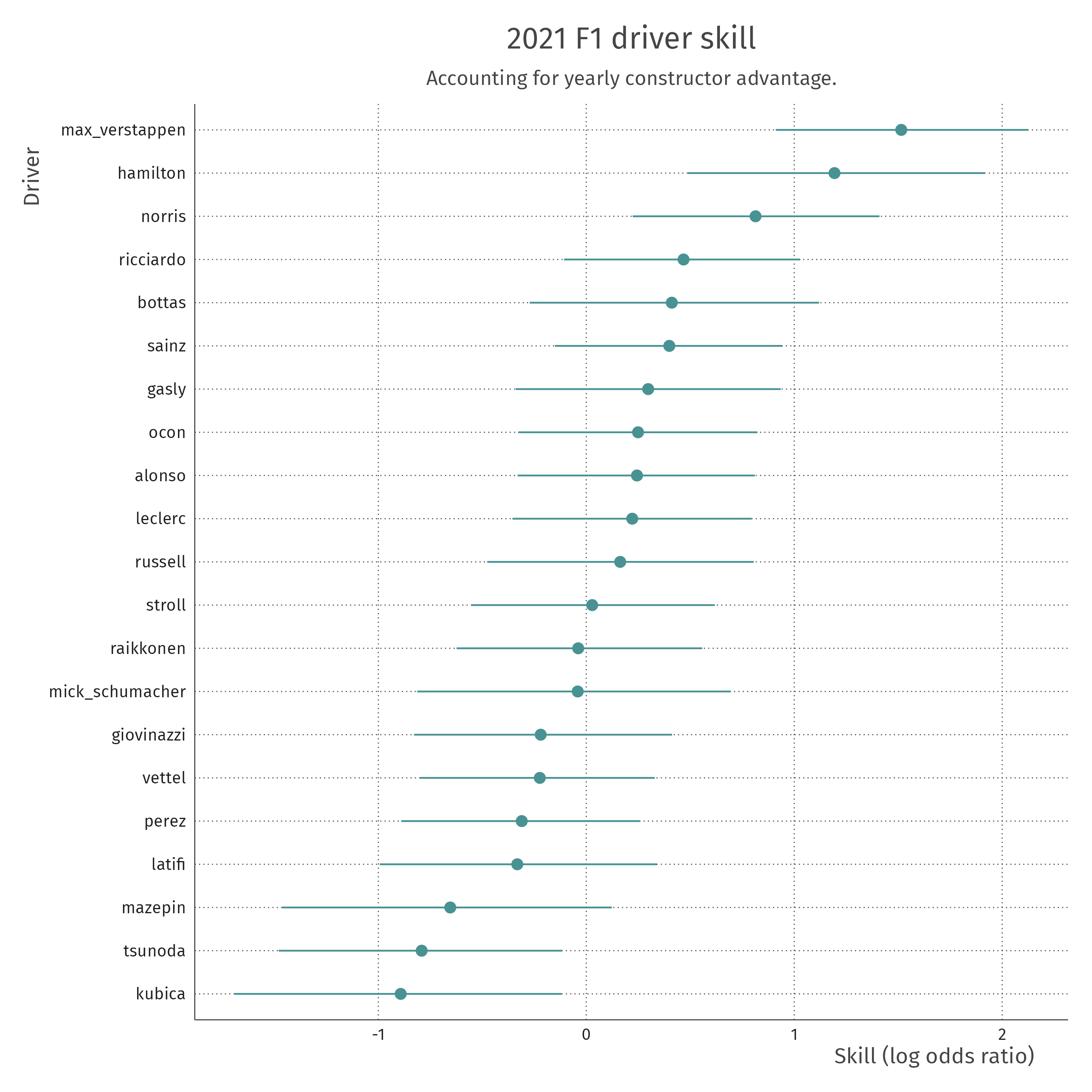}
    \caption{Driver skill in 2021 on the log odds ratio scale, accounting for yearly driver skill and constructor form. Error bars indicate 89\% credible interval.}
    \label{fig:driver_skill_2021}
\end{figure}

After estimation of the preferred model, it is possible answer the question of which driver is the most skilled, while taking into account constructor advantage, constructor form, as well as all parameter uncertainties. In order to produce a ranking, we obtained the posterior means and 89\% credible intervals \citep[see][]{mcelreath2018statistical} of $\theta_d + \theta_{ds}$ for the season 2021. These summaries are shown in Figure \ref{fig:driver_skill_2021}.

In Figure \ref{fig:driver_skill_2021} it is apparent that of the 2021 drivers, Hamilton and Verstappen are ranked as the most skilled driver. Note that this ranking comes directly from the model, which is estimated only on the hybrid-era data. Earlier performances by drivers such as Vettel (four-time world champion in the period 2010--2013) and Räikkönen (world champion in 2007) have not been taken into account, explaining their lower position on the ranking.

\begin{figure}[H]
    \centering
    \includegraphics[width=0.8\textwidth]{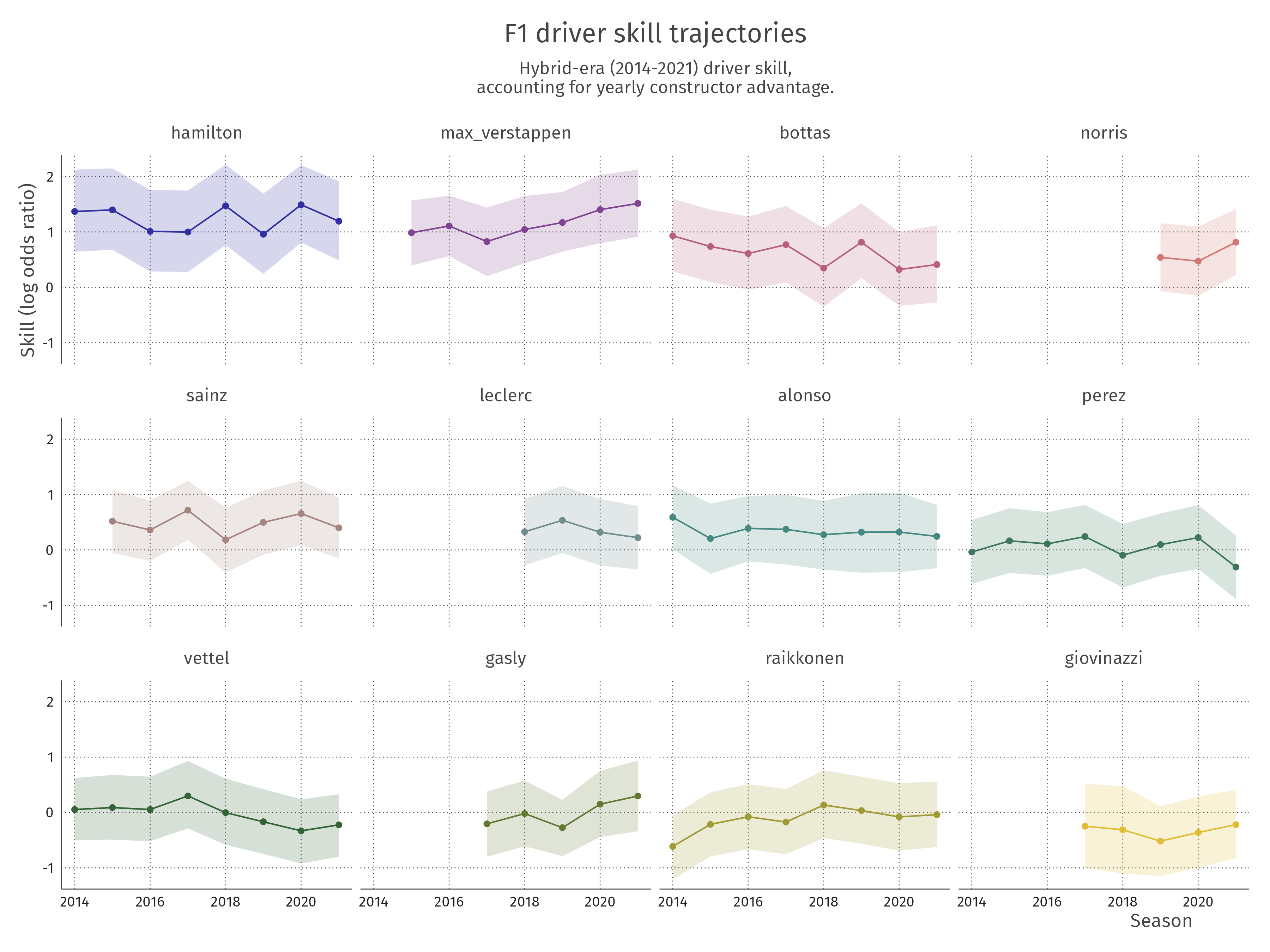}
    \caption{Driver skill trajectories for 12 drivers in the hybrid era on the log odds ratio scale. Ribbons indicate 89\% credible interval.}
    \label{fig:skill_trajectory}
\end{figure}

Because the model contains a yearly form parameter, we can also visualize the latent skill trajectories of several drivers throughout the hybrid era, with their credible intervals. The result of this visualization for 12 drivers of the 2021 season is shown in Figure \ref{fig:skill_trajectory}.

The figure shows that there are slight changes in skill across seasons. Drivers such as Verstappen, Norris, and Gasly tend to improve over years, on average, whereas Bottas displays a slight decline. Notably, Hamilton is consistently at the top, whereas Alonso and Sainz are consistently slightly above average in terms of their latent skill.

\subsection{Constructor advantage}
\begin{figure}[H]
    \centering
    \includegraphics[width=0.8\textwidth]{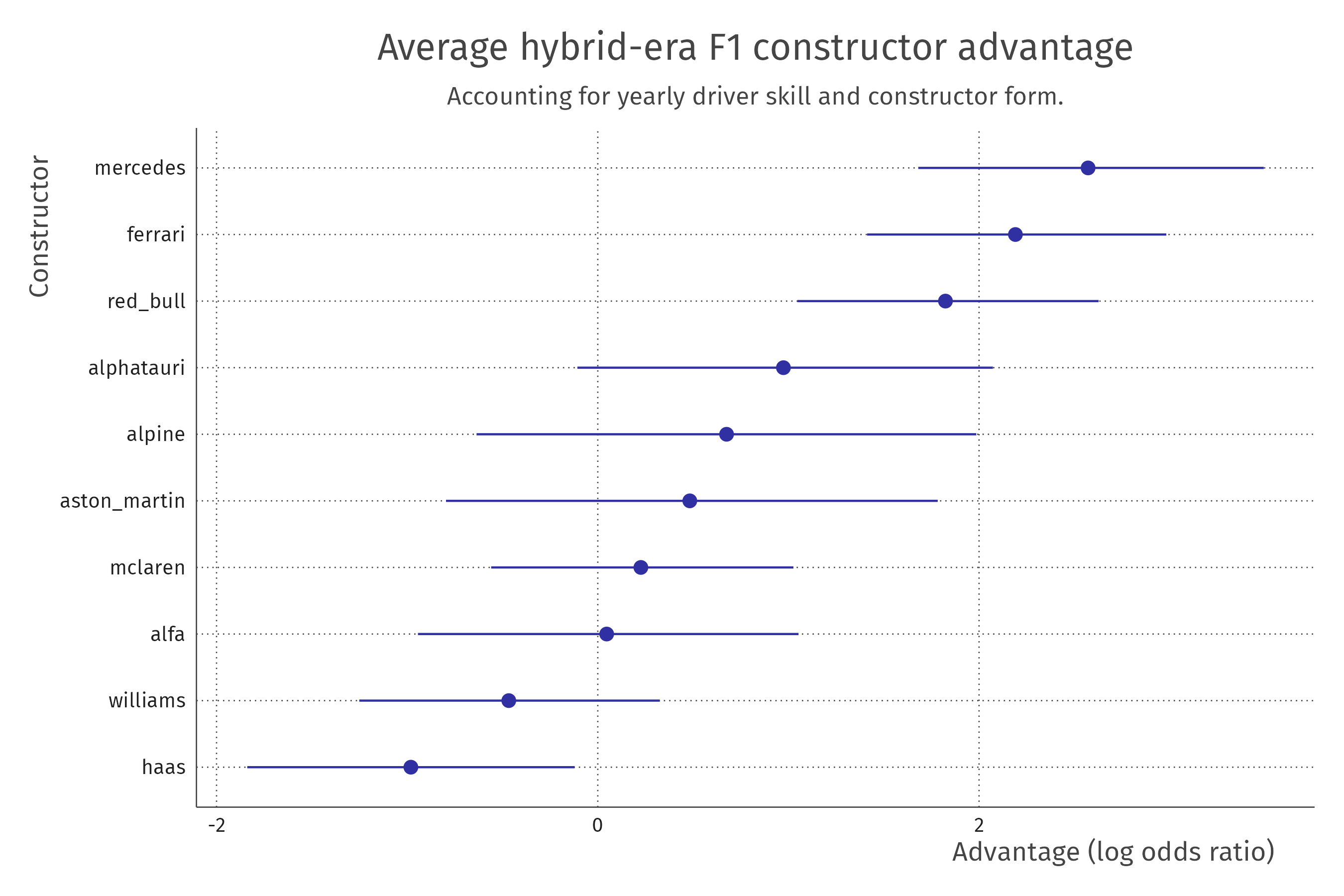}
    \caption{Long-run average constructor advantage on the log odds scale. Error bars indicate 89\% credible interval.}
    \label{fig:constructor_advantage_2021}
\end{figure}

For the constructors competing in 2021, we here investigate how much of an advantage their car yields. We do this by computing the posterior means and 89\% credible intervals of $\theta_t$ from the model. The result is shown in Figure \ref{fig:constructor_advantage_2021}.

One feature that becomes clear from this constructor advantage plot is the relative advantage of the \emph{big three} teams: Mercedes, Ferrari, and Red Bull have the largest budget and the most resources to spend on developing their car, which has resulted in these three teams excelling in the hybrid era. Another interesting feature in this plot is the large uncertainty around the teams that have competed in only a few seasons. For example, Alpine and Aston Martin were new teams in 2021, and therefore have not had a chance compete in many races, resulting in uncertainty around where it is placed in this constructor ranking. Again, these parameters need to be interpreted with care: they represent the average constructor advantage over the entire hybrid era.

\begin{figure}[H]
    \centering
    \includegraphics[width=0.8\textwidth]{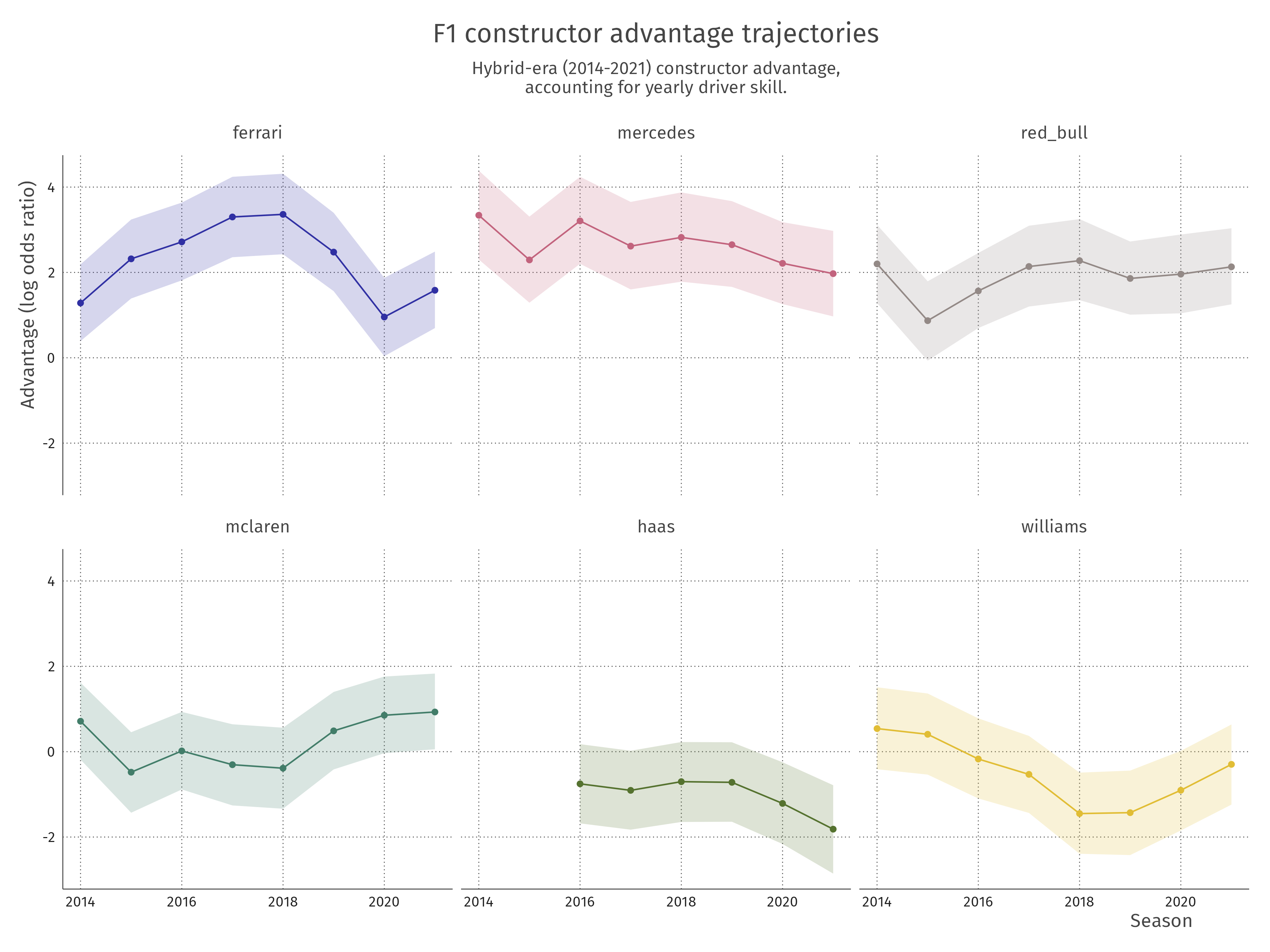}
    \caption{Yearly constructor advantage (summing constructor and constructor-year random effects) for Ferrari, Mercedes, Red Bull, McLaren, Haas, and Williams in the period 2014 -- 2020. Ribbons indicate 89\% credible interval.}
    \label{fig:advantage_trajectory}
\end{figure}

The last random intercept component is the constructor-year effects $\theta_{ts}$. These represent yearly constructor form, as a deviation from their long-term average advantage. In Figure \ref{fig:advantage_trajectory}, the yearly constructor advantage trajectories for a selection of teams is shown from 2014 to 2021.

One of the most striking features from this graph is Ferrari's drop in form from 2019 to 2020. There is a good explanation for this: after the 2019 season, there were allegations that Ferrari's engine in the previous year did not match league regulations. Ferrari subsequently reached a settlement with the Formula One Management \citep{f12020settlement}, which left the team with a relatively weak engine in the 2020 season. This change is aptly reflected in the constructor form parameter for Ferrari, but also to a lesser extent in that of McLaren, which profited from Ferrari's drop and ended up third in the constructor's championship that year.

\subsection{Season performance}
By combining driver skill and constructor advantage for each entrant in the 2021 season, we can create a posterior predictive simulation for the entire season. Here, we simulate finish positions (for the races where the entrant actually finished), which we then translate to points using a slightly simplified version of the 2021 points system (i.e., excluding fastest lap points and sprint race points). From this, we compute the expected average points per race over the whole 2021 season, as well as its 89\% credible interval. Then, we compare this to the realized average points for each entrant using the same simplified points system. The result of this is shown in Table \ref{tab:rank}.

\begin{table}[ht]
\centering
\begin{tabular}{rlcc}
  \hline
  Driver & Constructor & Expected points [89\% CI] & Observed points \\ 
  \hline
  Verstappen & Red Bull     & 15.30 [12.32, 18.00] & 18.00 \\
  Hamilton   & Mercedes     & 14.58 [11.45, 17.50] & \underline{17.59} \\
  Bottas     & Mercedes     &  8.85 [5.95, 11.73] &  9.82 \\
  Sainz      & Ferrari      &  8.79 [5.73, 11.91] &  7.45 \\
  Norris     & McLaren      &  7.07 [4.23, 10.18] &  7.23 \\
  Leclerc    & Ferrari      &  7.00 [4.18,  9.95] &  7.32 \\
  Perez      & Red Bull     &  6.75 [4.00,  9.73] &  8.59 \\
  Ricciardo  & McLaren      &  5.71 [3.09,  8.64] &  5.41 \\
  Gasly      & Alpha Tauri  &  5.40 [2.86,  8.14] &  5.14 \\
  Alonso     & Alpine       &  4.13 [1.95,  6.73] &  3.68 \\
  Ocon       & Alpine       &  3.95 [1.82,  6.50] &  3.50 \\
  Stroll     & Aston Martin &  2.61 [0.82,  4.86] &  1.55 \\
  Tsunoda    & Alpha Tauri  &  2.30 [0.68,  4.32] &  1.45 \\
  Vettel     & Aston Martin &  2.08 [0.55,  4.09] &  2.18 \\
  Raikkonen  & Alfa Romeo   &  1.93 [0.45,  3.82] &  0.45 \\
  Russell    & Williams     &  1.26 [0.14,  2.86] &  1.14 \\
  Giovinazzi & Alfa Romeo   &  1.95 [0.45,  3.86] & \underline{0.14} \\
  Latifi     & Williams     &  0.80 [0.00,  2.14] &  0.36 \\
  Schumacher & Haas         &  0.28 [0.00,  1.14] &  0.00 \\
  Mazepin    & Haas         &  0.12 [0.00,  0.73] &  0.00 \\
  \hline
\end{tabular}
\caption{For entrants in the 2021 season, the model's posterior expectations and 89\% credible interval for expected points per race (averaged over the 2021 season), and the realized 2021 average points per race (excluding points for fastest laps). Cases where the observed average points are outside of the 89\% CI are underlined in the last column.}
\label{tab:rank}
\end{table}

Generally, the expected and observed columns line up well in terms of credible interval coverage, although the model tends to underestimate points at the top and overestimate points at the bottom. This may be due to the regularizing effects of the multilevel model implementation, both on average performance and seasonal form.

\subsection{Relative contributions of drivers and constructors}

In order to investigate the contributions of drivers and constructors to the race results, we investigate the standard deviations of the random intercepts in the model. By investigating these standard deviations, we can make conclusions about which matters more in terms of race results: the driver or the car. The posteriors for the variation coefficients are shown in Table \ref{tab:ranef}.

\begin{table}[ht]
\centering
\begin{tabular}{rccccc}
  \hline
  Component & Symbol & Estimate & Est.Error & Lower & Upper \\ 
  \hline
  Constructor advantage & $\sigma_c$ & {1.63} & {0.34} & {1.14} & {2.27} \\ 
  Constructor form & $\sigma_{cs}$   & {0.73} & {0.10} & {0.57} & {0.91} \\ 
  Driver skill & $\sigma_{d}$        & {0.54} & {0.12} & {0.35} & {0.76} \\ 
  Driver form & $\sigma_{ds}$        & {0.35} & {0.06} & {0.24} & {0.46} \\
  \hline
\end{tabular}
\caption{Standard deviations ($\sigma$) for the random effects in the model. Lower and upper represent bounds of the 89\% credible intervals.}
\label{tab:ranef}
\end{table}

The standard deviation of the constructor is larger than that for the driver. This means that on average, the constructor has a larger impact on race results than the driver. Rephrasing this, on average the correlation in the outcome is stronger for two different drivers driving for the same team than for the same driver driving for different teams. This interpretation is also exemplified by the race results shown in Figure \ref{fig:prop}: Räikkönen driving for Alfa Romeo (2020) looks more like his teammate (Giovinazzi) than like Räikkönen driving for Ferrari (2018).

Quantifying the relative importance of long-term constructor advantage compared to driver skill is also possible directly from the numerical summaries. The posterior estimates for the variances are as follows: $\sigma^2_c \approx 2.65$, $\sigma^2_{cs} \approx 0.54$, $\sigma^2_d \approx 0.29$, and $\sigma^2_{ds} \approx 0.12$. Following the methodology of \citet[][\S 4.2]{bell2016formula}, this means that constructor effects account for around 88\% of the variance in the model (89\% CI [0.775, 0.945]), which is very similar to the aforementioned authors who reported 86\%.

\subsection{Counterfactual inference}
Using samples from the posterior distributions of the parameters, we can answer some counterfactual questions about the drivers in the model. An example question would be: ``According to the model, would Hamilton be expected to beat Räikkönen in a race in 2021 if Hamilton drove for Alfa Romeo and Räikkonen for Mercedes?''.

\begin{figure}[H]
    \centering
    \includegraphics[width=0.8\linewidth]{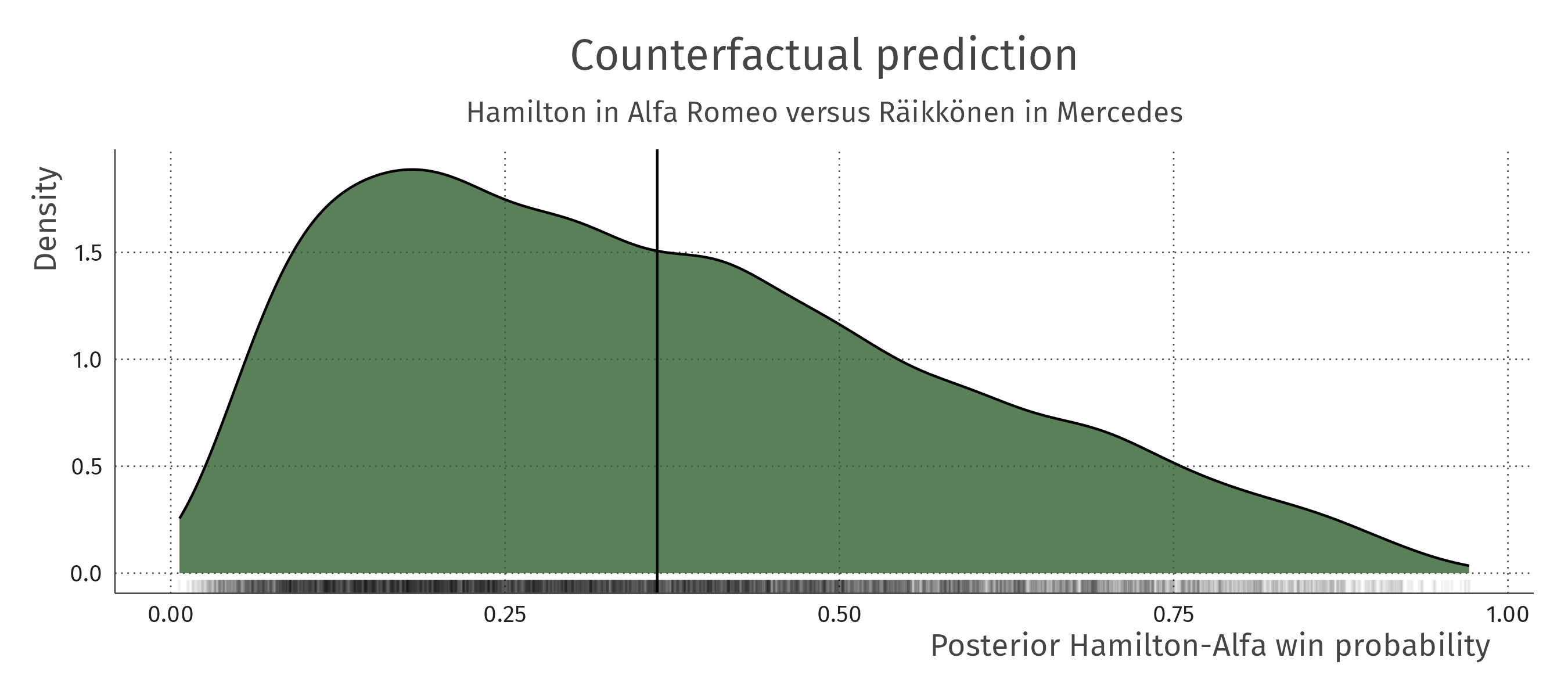}
    \caption{Posterior distribution of the win probability of Hamilton in an Alfa Romeo over Räikkönen in a Mercedes in a race during the 2021 season. The expected value of the distribution (vertical line) is 0.36, meaning that Räikkönen is expected to beat Hamilton in this scenario.}
    \label{fig:counter}
\end{figure}

We can answer this question by computing the posterior probability of Hamilton beating Räikkönen $\pi_{\rm ham>rai}$:
\begin{align}
    \theta_{\rm ham:alfa:2021} &= \theta_{\rm ham} + \theta_{\rm ham:2021} + \theta_{\rm alfa} + \theta_{\rm alfa:2021} \nonumber \\
    \theta_{\rm rai:merc:2021} &= \theta_{\rm rai} + \theta_{\rm rai:2021} + \theta_{\rm merc} + \theta_{\rm merc:2021} \nonumber \\
    \pi_{\rm ham>rai} &= \frac{\exp(\theta_{\rm ham:alfa:2021})}{\exp(\theta_{\rm ham:alfa:2021}) + \exp(\theta_{\rm rai:merc:2021})}
\end{align}
The posterior of $\pi_{\rm ham>rai}$ is shown in Figure \ref{fig:counter}. It is shown that Räikkönen is expected to beat Hamilton in this scenario ($E[\pi_{\rm ham>rai}] < 0.5$), with a clear degree of uncertainty.
Note that this counterfactual prediction is a way to summarise the model, meaning the same assumptions that accompany the model also accompany the counterfactual predictions. For example, for these predictions the data from before 2014 is irrelevant, and driver talent is independent of the constructor advantage.

\section{Discussion}
\label{sec:discussion}

In this paper, we used a Bayesian multilevel rank-ordered logit model for the rank results in Formula 1 races of the hybrid era (2014 -- 2021). After model development, comparison with leave-one-out cross-validation, and validation with posterior predictive checks, we have made inferences about the drivers and constructors competing in the 2021 season. In terms of the drivers, Hamilton and Verstappen are the best drivers. In terms of the constructors, the top three teams (Mercedes, Ferrari, and Red Bull) clearly outperform the rest on average in the hybrid era. Additionally, the model accurately represents changes in constructors' seasonal form, for example reproducing a drop in performance by Ferrari in 2020. Comparing driver contributions to team contributions, we have concluded that the car is more important than the driver when it comes to race results. Using our model and posterior sampling, it is possible to obtain answers of counterfactual questions as posterior distributions, which we have shown in the last part of Section \ref{sec:results}.


In terms of parameter interpretation, there is an interesting parallel between this model and Elo ratings \citep{elo1978rating} in chess. Both the Elo rating and our driver talent parameters can be transformed using an inverse logit to compare the relative strength of the competitors, and thus how likely it is that one competitor wins. The larger the difference between the ratings, the more certain it is that the competitor with the higher rating wins. We have shown such a comparison in a counterfactual situation in Section \ref{sec:results}. Note that this Bayesian hierarchical approach has been applied before to different sports \citep[e.g., in tennis;][]{ingram2019point}. However, in our model not only the athletes get ratings, but also the constructors, so comparisons can be made at this level as well. This approach could be used in other sports where multiple independent components contribute to competition results.


While this model does well at describing past data, for example closely reproducing the ranking of the 2021 season, it is probably not suitable for prediction. It uses very limited information (only the driver, constructor, and year) and for each year a specific effect needs to be estimated. Before a season starts, there is no data on that season, meaning that these year-effects are unavailable (even though they are important components of this model). For forecasting, approaches such as that of \citet{henderson2018comparison} may be more suitable relative to the baseline of bookmaker odds.

There are several areas where this model may be improved. One area is in team continuity: teams can officially change their name, when behind the scenes it is the same team, with the same long-run performance. For example, Alpha Tauri is a re-branding of the Italian Toro Rosso team, but it enters our dataset as a completely new team, with understandably large credible intervals around performance in 2020. By not accounting for team continuity across different team names, we had in total 17 different constructors in the dataset. On the other hand, team name changes often do go hand-in-hand with some structural changes, and it is hard to determine the extent to which this happens: where to draw the line between a ``re-brand'' and a new team?

While our model accurately represents driver and constructor performances in the seasons 2014 -- 2020, the data range could be expanded in order to provide results that better reflect the careers of certain drivers. For example, Räikkönen -- who is at the end of his career -- is in the bottom half in Figure \ref{fig:driver_skill_2021} and Table \ref{tab:rank}. Because the biggest part of Räikkönen's career is missing from the data, his 2007 world championship has not been appropriately taken into account. These errors propagate, perhaps overestimating the performance of Alfa Romeo (Räikkönen's team for that year) and underestimating the skill of Giovinazzi (whose only teammate has been Räikkonen).

\newpage
\bibliographystyle{apalike}
\bibliography{refs}
\newpage
\appendix
\appendixpage
\section{Sensitivity analysis for race finishing}
\label{app:finish}
In the main paper, we remove non-finishers from the data, estimating the model parameters data only from the races in which a certain competitor finished the race. This choice influences the interpretation of the parameters: the skill parameters exclude any measure of "reliability" because only finished races count. In this section, we perform sensitivity analysis for this choice, investigating the effects on the results of making a different choice here: 

\begin{enumerate}
    \item Not removing any results, i.e., ranking unfinished competitors based on race distance. 
    \item Removing only car-related non-finishes and ranking the driver-related non-finishes as normal
    \item Removing only driver-related non-finishes and ranking the car-related non-finishes as normal
    \item Only ranking finished competitors (i.e., original analysis from main paper)
\end{enumerate}

As mentioned in the main text in section \ref{sec:data}, yet another option is to include non-finishes in the model directly, resulting in a joint binomial model for finishing and rank model for ranking conditional on finishing. This approach is outside the scope of this sensitivity analysis. 

In the data, it is difficult to accurately ascertain which results belong to car-related and driver-related errors: there are 51 different status types other than finishing the race; some are ambiguous with respect to who caused them, e.g., ``Excluded'', ``Technical'', ``Damage''. For the purpose of this sensitivity analysis, we have made the following selection: 

\begin{itemize}
    \item[\textbf{Driver-related}: ] Collision, Disqualified, Withdrew, Retired, Accident, Collision damage, Spun off, Excluded, Illness
    \item[\textbf{Car-related}: ]  ERS, Oil pressure, Engine, Technical, Gearbox, Electrical, Power Unit, Brakes, Clutch, Exhaust, Mechanical, Turbo, Rear wing, Drivetrain, Suspension, Oil leak, Water leak, Water pressure, Electronics, Transmission, Wheel, Power loss, Fuel system, Front wing, Tyre, Throttle, Brake duct, Hydraulics, Battery, Puncture, Overheating, Wheel nut, Vibrations, Driveshaft, Fuel pressure, Seat, Spark plugs, Steering, Damage, Out of fuel, Debris, Radiator
\end{itemize}

\begin{figure}
    \centering
    \includegraphics[width=\linewidth]{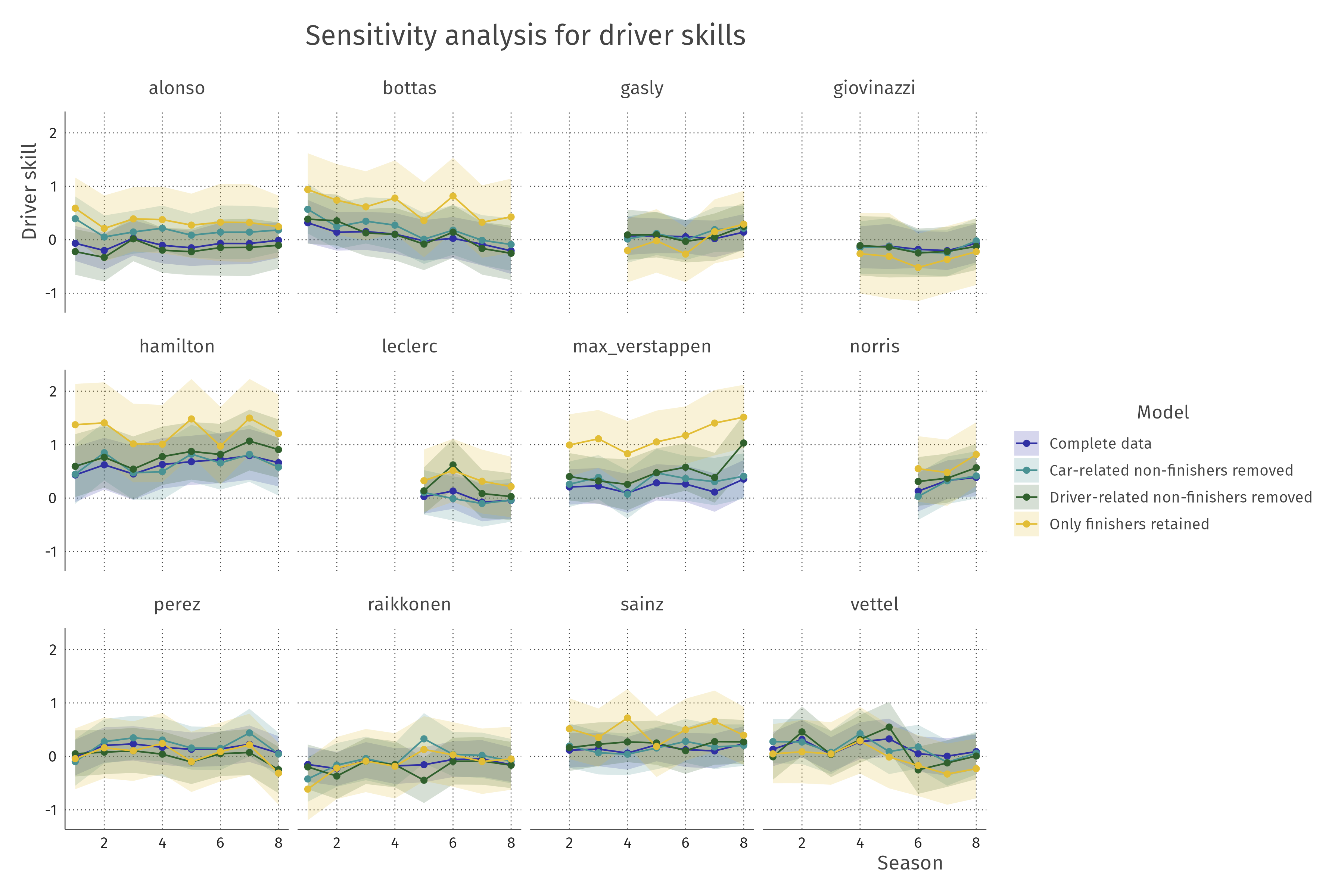}
    \caption{Sensitivity analysis for driver skills inferences.}
    \label{fig:sensdriv}
\end{figure}

\begin{figure}
    \centering
    \includegraphics[width=\linewidth]{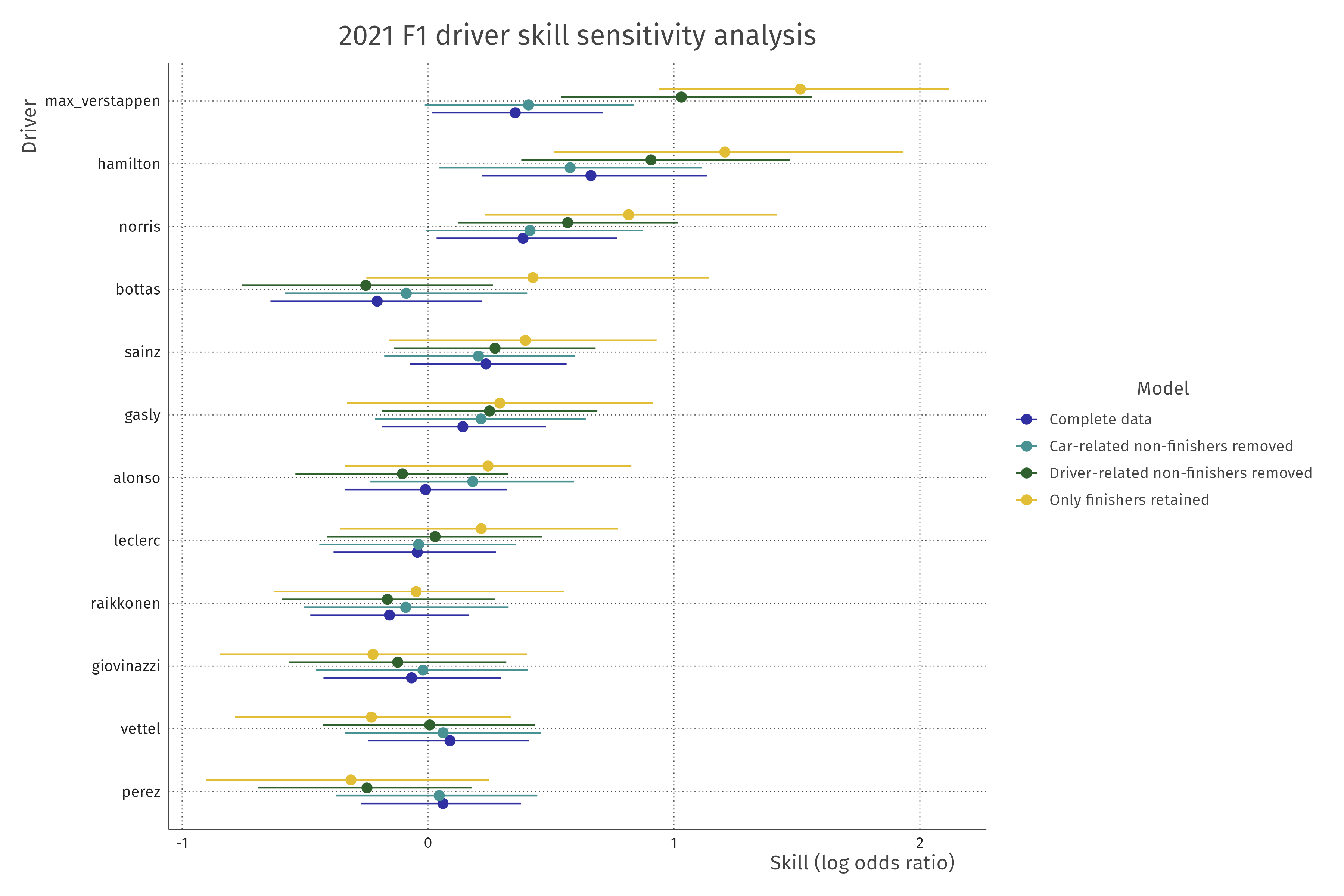}
    \caption{Sensitivity analysis for driver skills inferences for the 2021 season.}
    \label{fig:sensdriv2021}
\end{figure}

\begin{figure}
    \centering
    \includegraphics[width=\linewidth]{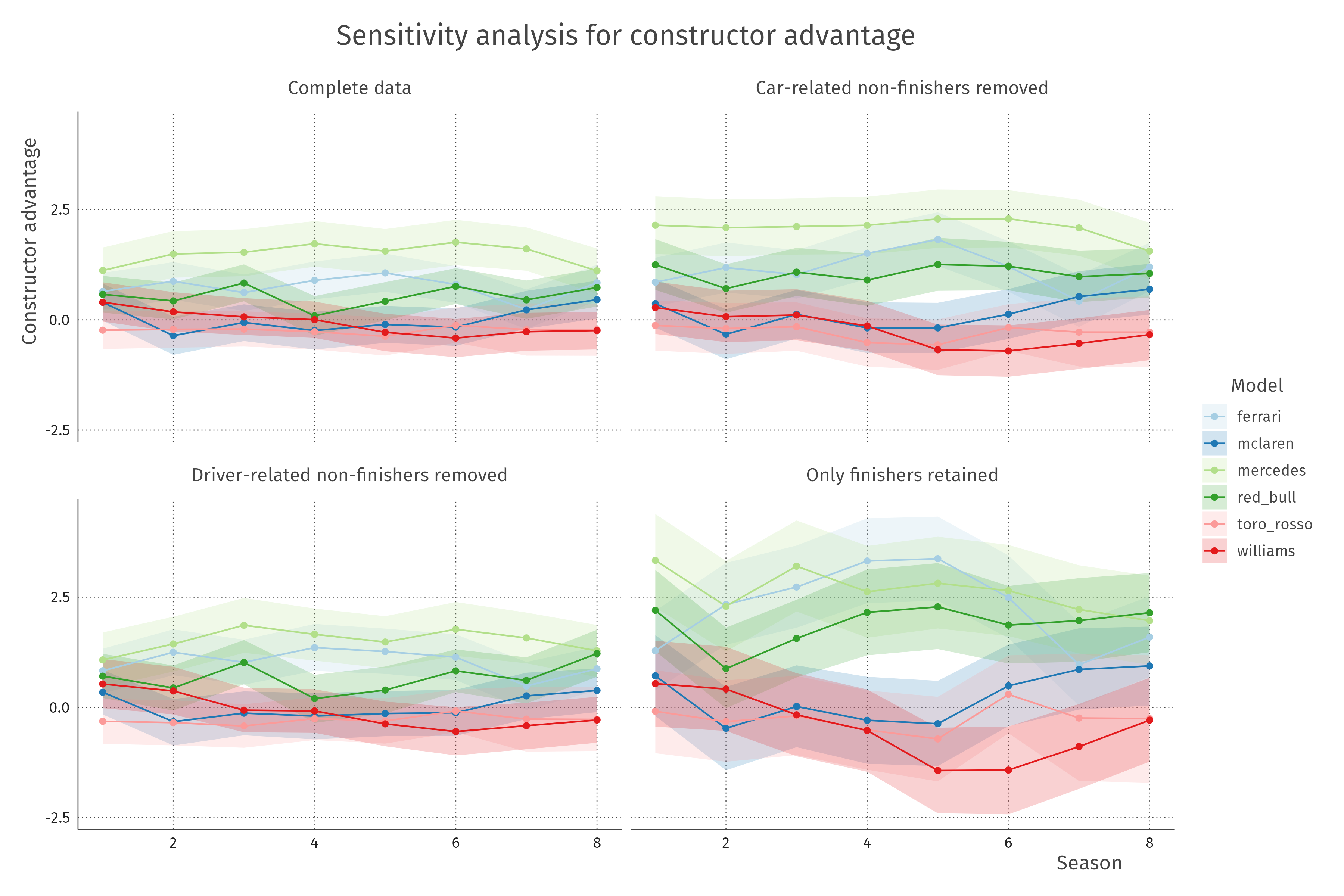}
    \caption{Sensitivity analysis for constructor advantage inferences.}
    \label{fig:sensteam}
\end{figure}

The results of the sensitivity analysis are shown graphically in Figures \ref{fig:sensdriv}, \ref{fig:sensdriv2021} (driver skills) and \ref{fig:sensteam} (constructor advantage). A general pattern in comparing the main paper model ("only finishers retained") to the other three models is the smaller variance of these random effect components; when we include non-finishes in the ranking data, this ranking becomes more noisy so the latent skill and advantage parameters will be closer together (and closer to 0). This is especially visible in Figure \ref{fig:sensteam}, which was faceted by the model variable for this purpose.

For the drivers, generally the skill patterns over time (Figure \ref{fig:sensdriv}) are similar across different models, meaning that skill increases and decreases are captured somewhat similarly for the different models. Looking at the skill level in 2021, we see that a notable outlier is Bottas, who drops dramatically in skill level relative to his surrounding drivers once non-finishes are included in the model. It is unclear what the mechanism is behind this drop, as Bottas has not had much more non-finishes than other drivers (4 per season, whereas the average is 3).

In summary, the results are somewhat sensitive to the choice of what to include in the model. The conclusions in the main paper should be interpreted carefully; we once again make clear that the interpretation of the skill parameters only includes \emph{finished} races, thus do not contain any component of reliability of the car or carefulness of the driver. For example, Pastor Maldonado --- who was notoriously crash-prone --- is ranked $6^{th}$ worst if we include all data in the model, but a much better $19^{th}$ of 38 drivers if we only look at finished races. 

\newpage

\section{Extended model definition}
\label{app:model}
In this section, we extend the model definition in Section \ref{sec:model} in the main paper into a fully specified generative model. Following \citet{glickman2015stochastic}, the Rank-Ordered Logit (ROL) distribution implies that the skills for each competitor in the $r^{th}$ race ($\boldsymbol{\vartheta}_{r}$) generate a latent \emph{performance} vector $\boldsymbol{z}_r$ following independent standard extreme value (Gumbel) distributions. The resulting rank-ordering $\boldsymbol{y}_{r}$ is the rank-ordering of these latent performance values. By replacing RankOrderedLogit by this generative process, the full model can be specified as follows:
\begin{align}
    \boldsymbol{y}_{r} &= {\rm rank}(\boldsymbol{z}_r) \nonumber \\
    \boldsymbol{z}_r &= \{z_c \,| \,c \in \mathcal{C}_r\} \nonumber \\
    z_c &\sim {\rm Gumbel}(\vartheta_c) \nonumber \\
    \vartheta_{c} = \theta_{dts} &= \theta_d + \theta_{ds} + \theta_t + \theta_{ts} \nonumber \\
    \theta_d &\sim \mathcal{N}(0, \sigma^2_d) \nonumber \\
    \theta_{ds} &\sim \mathcal{N}(0, \sigma^2_{ds}) \nonumber \\
    \theta_t &\sim \mathcal{N}(0, \sigma^2_t) \nonumber \\
    \theta_{ts} &\sim \mathcal{N}(0, \sigma^2_{ts})
    \label{eq:appmodel}
\end{align}
where we make use of the standard Gumbel distribution with only a location parameter $\mu$ and no scale:
\begin{equation}
    f(x; \mu) = \exp\left(x - \mu - \exp(x - \mu)\right)
\end{equation}

The likelihood for the model in equation \ref{eq:appmodel} can be simplified for implementation \citep[][eq. 4]{glickman2015stochastic}. Assuming, without loss of generality, that the set of competitors for a specific race $\mathcal{C}_r$ is already ordered according to the race result $\boldsymbol{y}_{r}$, we obtain the following likelihood:
\begin{equation}
    p(\boldsymbol{y}_{r} | \boldsymbol{\vartheta}_r) = p(z_1 > z_2 > ... > z_{m_r} | \boldsymbol{\vartheta}_r) = \prod_{i = 1}^{m_r-1} \frac{\exp{(\vartheta_i)}}{\sum_{j = i}^{m_r} \exp{(\vartheta_j)}}
\end{equation}
where $m_r$ is the number of competitors $|\mathcal{C}_r|$. 

\newpage

\section{Traceplot}
\label{app:trace}

\begin{figure}[H]
    \centering
    \includegraphics[width=0.8\textwidth]{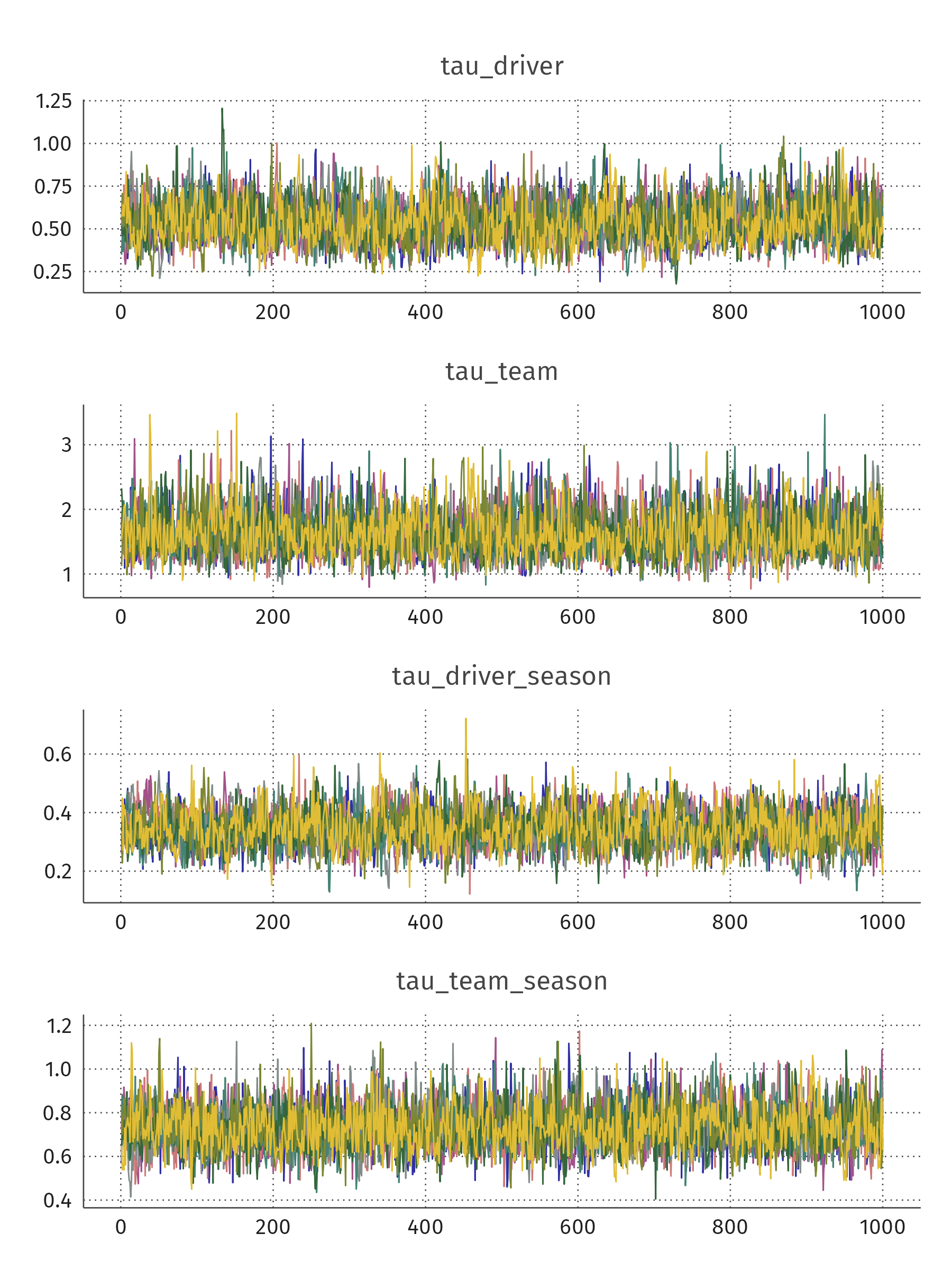}
    \caption{Monte carlo markov chain visualisation for the main model parameters, indicating satisfactory convergence for all shown parameters.}
    \label{fig:chains}
\end{figure}

Figure \ref{fig:chains} shows the traceplot for the main standard deviation parameters of the rank-ordered logit model. From top to bottom, the standard deviations for the four random effects $\beta_d$, $\beta_{ds}$, $\beta_c$, and $\beta_{cs}$ are shown.
All traceplots for the four different chains overlap as expected from an appropriately converged model.

\newpage

\section{Comparing different dynamic skill parameter implementations}
\label{app:dynamic}

In the paper, the model does not take time into account explicitly, but it estimates i.i.d. seasonal form parameters per driver and constructor. Here, this approach is briefly compared to explicitly using a latent multilevel AR(1) implementation (\texttt{auto}), and to a more parsimonious multilevel intercept + slope model (\texttt{slope}) for the latent skill parameters. For the auto-regressive model, we used a cluster mean centered parameterization as in \citet{hamaker2015center}.

Because these models have the same likelihood form (the rank-ordered logit likelihood), we can compare them using efficient leave-one-out cross-validation:

\begin{table}[H]
\centering
\centering
\begin{tabular}{rcccc}
  \hline
  & $ELPD$ & $SE_{ELPD}$ & $\Delta$ & $SE_{\Delta}$ \\ 
  \hline
  Auto  & -3992.7 & 83.9 &       &      \\ 
  Rank  & -3993.8 & 83.9 & -1.1  & 1.7  \\ 
  Slope & -4042.9 & 81.6 & -50.2 & 18.4 \\ 
  \hline
\end{tabular}
\end{table}

The simple slope model is quite clearly worse than the other two implementations; it is not able to capture the skills properly. For example, looking at Giovinazzi and Räikkönen in the plot below shows that they are estimated to be top drivers in the 2021 season, which is unlikely given their results.

\begin{figure}[H]
    \centering
    \includegraphics[width=0.85\linewidth]{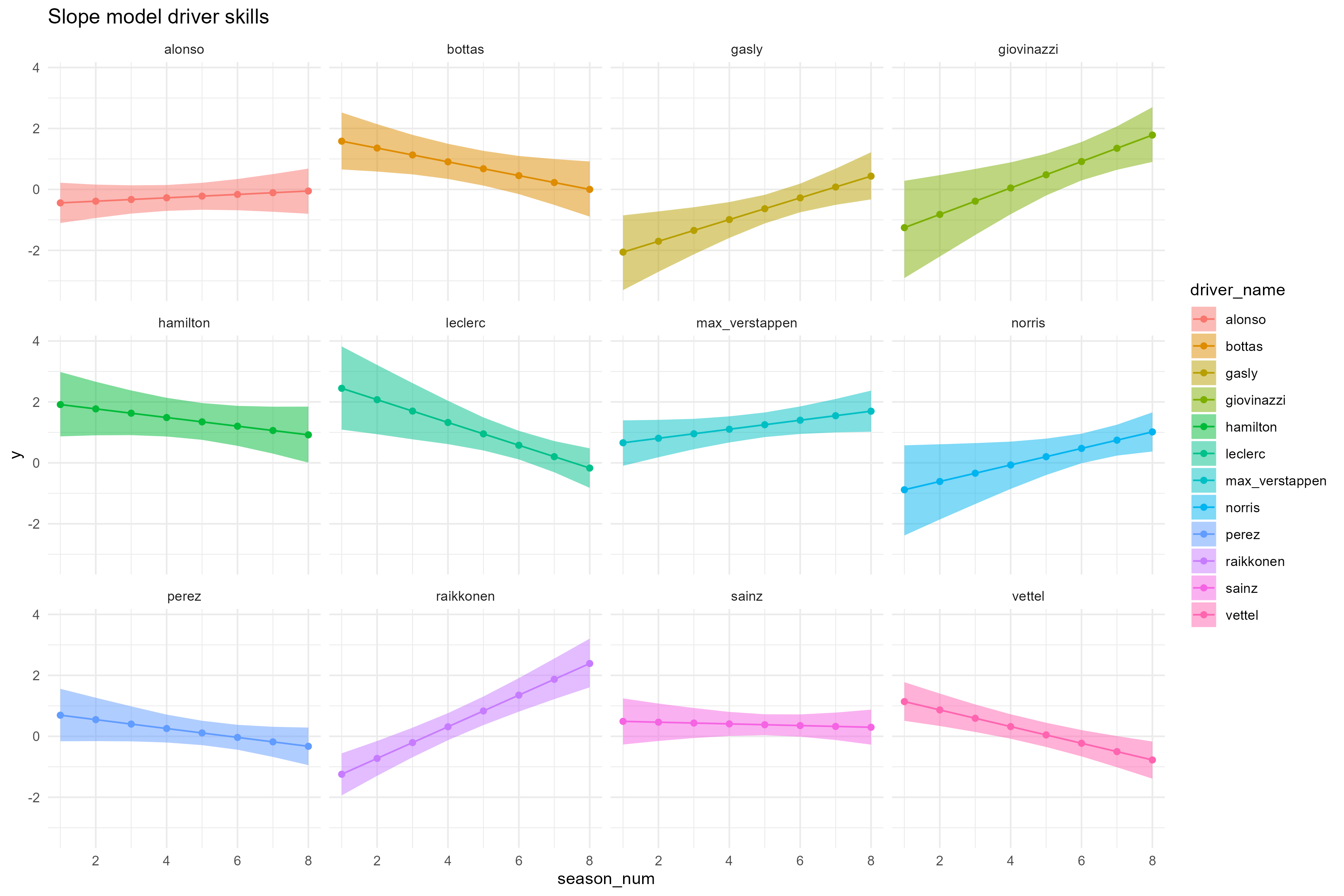}
    \caption{Posterior distribution of driver skill parameters over seasons for the slope model. Ribbons indicate 89\% credible interval.}
    \label{fig:drivslope}
\end{figure}

The AR(1) and default rank model are very similar in terms of LOOCV performance. Looking at the estimated skill trajectories, it becomes clear why this is the case:

\begin{figure}[H]
    \centering
    \includegraphics[width=0.85\linewidth]{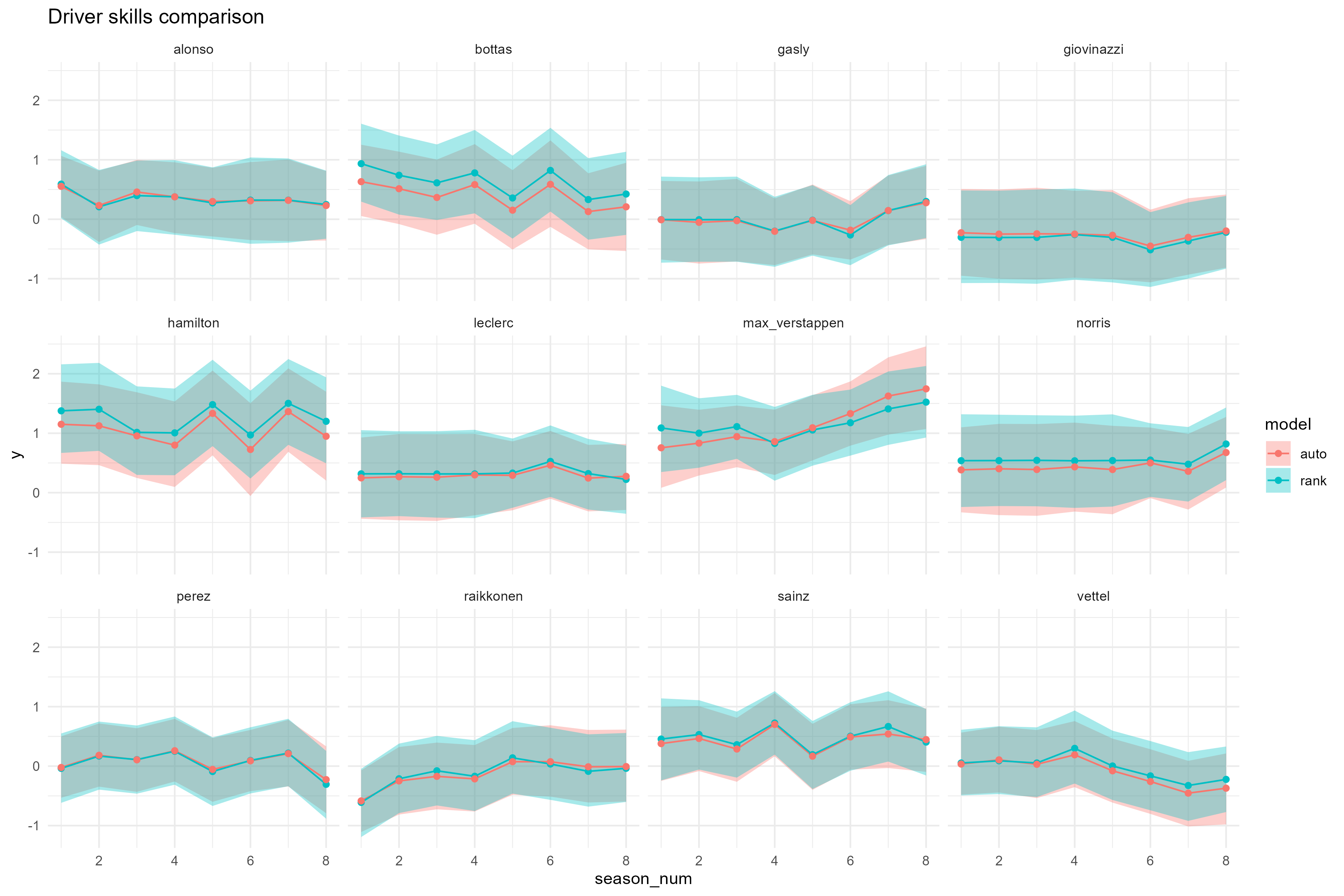}
    \caption{Posterior distribution of driver skill parameters over seasons for the multilevel and AR(1) implementations. Ribbons indicate 89\% credible interval.}
    \label{fig:drivar}
\end{figure}

The driver skills have very similar estimates. The same holds for the constructor advantage over time:

\begin{figure}[H]
    \centering
    \includegraphics[width=0.85\linewidth]{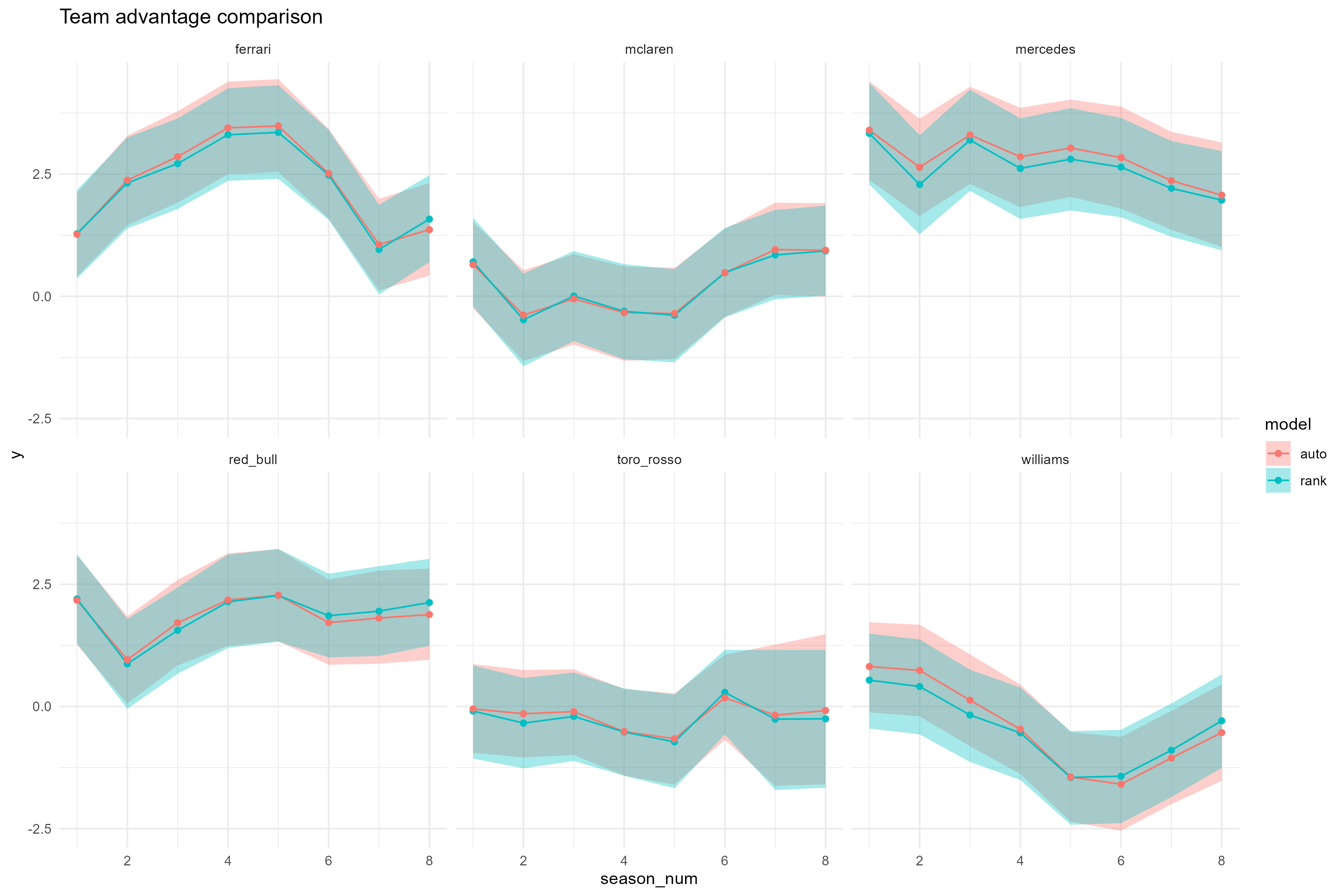}
    \caption{Posterior distribution of team advantage parameters over seasons for the multilevel and AR(1) implementations. Ribbons indicate 89\% credible interval.}
    \label{fig:teamar}
\end{figure}

Based on this comparison, the decision was made to work with the more interpretable basic multilevel model in the main paper.

\end{document}